\documentclass[%
reprint,
superscriptaddress,
nofootinbib,
 amsmath,amssymb,
 aps,
prd,
algorithm,
]{revtex4-2}

\usepackage{graphicx}
\usepackage{dcolumn}
\usepackage{bm}
\usepackage{xcolor}
\usepackage{xspace}
\usepackage{printlen}
\usepackage{ulem}
\usepackage{hyperref}
\usepackage{url}
\usepackage{microtype}
\usepackage{orcidlink}

\usepackage{algorithm}
\usepackage{algpseudocode}	
\usepackage{booktabs} 
\usepackage{multirow}
\usepackage{longtable}
\usepackage{tabularx}

\newcommand{\furax}{\texttt{FURAX}\xspace}
\newcommand{\toast}{\texttt{toast}\xspace}
\newcommand{\sotodlib}{\texttt{sotodlib}\xspace}
\newcommand{\cadre}{\texttt{CADRE}\xspace}
\newcommand{\namaster}{\texttt{NaMaster}\xspace}

\newcommand{\ItoP}{$I \rightarrow P$\xspace}

\newcommand{\vv}{\mathbf}       
\newcommand{\mat}{\mathbf}             

\newcolumntype{P}[1]{>{\centering\arraybackslash}p{#1}}     

\begin{document}


\title{
Robust CMB polarisation mapmaking with a rotating half-wave plate
}

\author{Wuhyun Sohn\,\orcidlink{0000-0002-6039-8247}} \email{sohn@apc.in2p3.fr}
\affiliation{Université Paris Cité, CNRS, Astroparticule et Cosmologie, F-75013 Paris, France}

\author{Simon Biquard\,\orcidlink{0000-0002-1493-2963}}
\email{simon.biquard@manchester.ac.uk}
\affiliation{Jodrell Bank Centre for Astrophysics, University of Manchester, Oxford Road, Manchester, England M13 9PL, United Kingdom}
\affiliation{Université Paris Cité, CNRS, Astroparticule et Cosmologie, F-75013 Paris, France}

\author{Josquin Errard\,\orcidlink{0000-0002-1419-0031}}
\email{josquin@apc.in2p3.fr}
\affiliation{Université Paris Cité, CNRS, Astroparticule et Cosmologie, F-75013 Paris, France}

\author{Radek Stompor\,\orcidlink{0000-0002-9777-3813}}
\email{radek.stompor@in2p3.fr}
\affiliation{Université Paris Cité, CNRS, Astroparticule et Cosmologie, F-75013 Paris, France}

\date{\today}
\hspace{10pt}

\begin{abstract}

We present a novel mapmaking method for obtaining unbiased estimates of CMB polarisation, tailored to modern CMB experiments with a rotating half-wave plate. These experiments are exposed to strong unpolarised contaminant sources, such as atmospheric emission and ground pickup, which can be several orders of magnitude stronger than the sky signal. Our mapmaker mitigates these systematic effects by marginalising over all signals that vary slowly compared to the timescale of a polarimeter's angle rotation on the sky, while recovering high-fidelity polarisation maps. When the variability timescales of the unpolarised signals exceed a quarter of the half-wave plate rotation period, the method can produce maps with nearly optimal noise levels and minimal contamination. Furthermore, if the half-wave plate rotation period is sufficiently short relative to the beam-scale crossing time, the method efficiently mitigates the sky intensity-to-polarisation leakage. This mapmaker, named the Polarisation-Optimised Map-Making Estimator (POMME), is implemented within the open-source FURAX package and is ready for application to upcoming ground-based CMB surveys.

\end{abstract}

\maketitle


\section{Introduction}

Observations of the Cosmic Microwave Background (CMB) have played a central role in establishing the standard cosmological model while providing new, important perspectives on fundamental physics. Measurements of CMB polarisation probe the physics of recombination, reionisation, and inflation. In particular, detecting primordial $B$-modes would have significant implications for early universe physics. Extracting physical insights from CMB sky maps requires mapmaking methods that deliver high-fidelity estimates of the Stokes parameters $Q$ and $U$ from increasingly large and complex time-ordered data (TOD) sets.

Ground-based CMB experiments, capable of deploying a large number of detectors and operating for extended periods of time, provide one of the most promising opportunities to achieve these science objectives. However, they also face numerous challenges. These include atmospheric emission, ground pickup (scan-synchronous signals), and instrumental systematics, all of which can exceed the sky polarisation signal by several orders of magnitude, e.g.~\cite{POLARBEAR2017, CLASS2025}. These predominantly unpolarised contaminants not only lead to loss of sensitivity (by increasing the optical loading of detectors), but also may be partially converted by instrumental systematics into a spurious polarised signal if left unmitigated\textemdash commonly referred to as \ItoP\ leakage~\cite{Hu2003, Shimon2008}.

Modern experiments employ specific hardware solutions for effective and robust measurements of the polarisation signal. For example, it is common to install an orthogonal pair of antennas at each pixel of the focal plane. The co-located pair of detectors probes the same total intensity but orthogonal polarisation directions, so that the differenced signal directly accesses polarisation with minimal total intensity contributions. Rotating half-wave plates (HWPs) have become increasingly popular with modern ground-based CMB experiments. A HWP modulates the linear polarisation signal into a well-defined frequency band centred around $4 f_{\rm HWP}$. This effectively shifts sky polarisation to higher frequencies in the TOD, away from frequencies dominated by unpolarised atmospheric emission and many systematic effects~\cite{Kusaka2014,Bryan2010}.

Advances in CMB instruments lead to new features of observational data, which in turn modify underlying mathematical models, providing new opportunities for data analysis methods to adapt. The first major compression step in any CMB data analysis pipeline is mapmaking, where large volumes of TOD are reduced by a factor of $\mathcal{O}(10^5-10^7)$ into the estimates of the sky signal on a pixelised map. The full information concerning the instrument operations, such as sky pointing and polarimeter angle, is usually assumed to be available and ideally should be fully capitalised on in order to ensure the production of high-fidelity maps of the sky. This should include an exploitation of specialised hardware solutions, such as antenna pairs and/or fast polarisation modulators. 

Mapmaking for CMB experiments has been studied extensively over nearly three decades~\cite{JanssenGulkis1992}, resulting in a plethora of methods and algorithms. Binned mapmakers provide fast and unbiased estimates under simplifying assumptions about the noise, but are generally sub-optimal in the presence of correlated noise. Maximum-likelihood (ML) approaches account for time-domain correlations through a non-diagonal noise covariance model, often approximated as Toeplitz~\cite{Tegmark1997, Stompor2001, Dore2001, deGasperis2005, Patanchon2008, act_mapmaking_2013}, but can be computationally demanding for next-generation data volumes~\cite{Cantalupo2010,ElBouhargani2022}. Template-based mapmakers, including destripers~\cite{Delabrouille1998, Revenu2000, Maino2002, Poutanen2004, Tristram2011}, introduce explicit models for low-frequency contaminants and marginalise over their amplitudes, providing an attractive compromise between robustness and computational cost.

These general methods can be adapted to the specific hardware solution and experimental approaches~\cite{Tristram2011, Kusaka2014, act_dr6_mapmaking_2025}. For instance, pair-differencing mapmaking approaches based on explicit differencing of simultaneous measurements from the two detectors of each orthogonal antenna pair have been developed, studied, and applied successfully to real data~\cite{BICEP22014, Polarbear2014, Crites2015}, producing high-quality estimates of the polarised sky signal. These require a careful assessment of all systematic effects that may affect the signals detected by two detectors differently. Moreover, they can only be applied to the antenna pairs for which both detectors work sufficiently well. Despite these limitations, it has been argued recently~\cite{biquard2025} that the pair-differencing combined with the continuously rotating HWP can be very efficient in suppressing many such effects and provides a robust and effective mapmaking tool for experiments featuring both these hardware solutions.

In this work, we introduce and study an alternative approach for producing sky polarisation maps. It does not explicitly assume the presence of orthogonal antenna pairs or make assumptions about the characteristics of different detectors. Instead, it suppresses all slowly varying unpolarised signals from detector timestreams separately, capitalising on the presence of the rotating HWP. We refer to this method as the ``\emph{Polarisation-Optimised Map-Making Estimator}'' (POMME), a template-based mapmaking method tailored to HWP-modulated CMB experiments. The core idea is simple: we marginalise over all signals that vary slowly in time by introducing piecewise-constant templates over intervals of duration $\tau$. This method can be thought of as a destriper~\cite{Keihanen2010} but with baseline lengths on the order of pixel-crossing times. The ``time-domain demodulation'' method of Ref.~\cite{manchesterDemodulation} also employs time intervals of fixed size but does so to separate, i.e., \textit{demodulate}, the Stokes parameters locally, which are subsequently binned in pixels on the sky. This is in contrast with our approach, where the intervals define the low-frequency modes which are then marginalised over in an explicitly unbiased manner.

We demonstrate that this procedure effectively mitigates low-frequency modes in the TOD and strongly suppresses atmospheric and scan-synchronous contamination. Since the cosmological polarisation signal is modulated to higher frequencies by the HWP, it remains largely unaffected by this deprojection. The method is computationally efficient, scales linearly with the data volume in its deprojection step, and yields unbiased polarisation maps under mild assumptions.

We characterise the performance and statistical properties of the estimator, derive analytic expressions for its noise covariance in simplified settings, and quantify the information loss associated with template marginalisation. We show that the choice of the interval length $\tau$ can be tuned to minimise noise degradation while maintaining strong suppression of low-frequency systematics. Through realistic simulations including atmospheric emission and scan-synchronous signals, we demonstrate that POMME significantly mitigates $I \rightarrow P$ leakage and delivers robust polarisation maps in regimes where conventional binned mapmakers fail. POMME is implemented within the FURAX framework\footnote{\url{https://github.com/CMBSciPol/furax}} and is designed for upcoming ground-based CMB surveys that employ rotating HWPs.

The remainder of this paper is organised as follows. In Section~\ref{sec:formalism} we present the formalism of template mapmaking and introduce the POMME estimator. Section~\ref{sec:implementation} describes the algorithmic implementation and simulation setup used for validation. The results on bias, noise properties, and robustness to atmospheric contamination are presented in Section~\ref{sec:results}. We conclude with a discussion of limitations and future extensions in Section~\ref{sec:discussion}.

\section{POMME formalism} \label{sec:formalism}

\subsection{Template mapmaking and deprojection}

For a ground-based CMB experiment with a rotating half-wave plate, our data model is given by~\cite{Johnson_2007} 
\begin{eqnarray}
    d_{i}(t) &=& I(t) + \cos(4\varphi(t)-2\alpha_i(t))Q(t) + \nonumber\\ &&\sin(4\varphi(t)-2\alpha_i(t))U(t) + T_i(t) + n (t), \label{eqn:data_model}
\end{eqnarray}
where stokes components $I$, $Q$, and $U$ are modulated by a half-wave plate of angle $\varphi$ and detector angle $\alpha_i$. Other contributions include non-sky-stationary signals $T_i$ and Gaussian noise $n$. For ground-based experiments, the former include, for instance, the unpolarised part of the atmospheric emission and ground-pickup. We note that while we assumed here an idealised model for the polarisation modulator, it is straightforward to extend the formalism to include more advanced and realistic descriptions, e.g.~\cite{ema2026}.

With discrete time samples and sky pixelisation, the data model in Eq.~\eqref{eqn:data_model} can be written in a general matrix form given by
\begin{align}
    \vv{d} = \mat{P}\vv{s} + \mat{T} \vv{x} + \vv{n}.   \label{eqn:template_mapmaking_model}
\end{align}
Note that all measurements from individual detectors have been stacked into a single long vector $\vv{d}$. The contributions from the celestial sky comprise three Stokes components: $\vv{s}\equiv(\vv{s}_I, \vv{s}_Q, \vv{s}_U)$. The pointing matrix $\mat{P}$ takes $\vv{s}$, projects their values to time samples based on detector pointing, rotates their Stokes $Q/U$ parameters using detector and half-wave plate angles, before finally combining them into bolometric measurements through a polariser. The templates $\mat{T}$ model various, unpolarised non-sky contributions, with the amplitudes $\vv{x}$ generally unknown. The noise is often assumed to be Gaussian: $\vv{n}\sim\mathcal{N}(\vv{0},N)$. A general, unbiased sky signal estimator is then given by
\begin{align}
    \hat{\vv{s}} &\equiv (\mat{P}^\top \mat{F}_\mat{T} \mat{P})^{-1} \mat{P}^\top \mat{F}_\mat{T} \vv{d},
    \label{eqn:template_mapmaking} 
\end{align}
where,
\begin{align} 
    \mat{F}_\mat{T} &\equiv \mat{W} \mat{D}_\mat{T} \equiv \mat{W}\left[ \mat{I}  - \mat{T}(\mat{T}^\top \mat{W} \mat{T})^{-1} \mat{T}^\top \mat{W} \right], \label{eqn:template_mapmaking_FT}
\end{align} 
if no priors are imposed on the template amplitudes~\cite{poletti2017}, or,
\begin{align}
    \mat{F}_\mat{T} &\equiv \mat{W}\left[ \mat{I} - \mat{T}(\mat{\Sigma}^{-1}_\mat{T} + \mat{T}^\top \mat{W} \mat{T})^{-1} \mat{T}^\top \mat{W} \right],\label{eqn:template_mapmaking_FT_w_priors}
\end{align}
if such priors are present~\cite{biquard2025phd}. Here, $\mat{W}$ denotes an arbitrary, symmetric and positive-definite weight matrix.

If the template term $\mat{T}\vv{x}$ can be neglected, and the weight matrix $\mat{W}$ is assumed to be diagonal, then the estimator is equivalent to the \textit{binned} mapmaker. Although diagonal weights significantly reduce the computational cost, they are a poor approximation to realistic noise and lead to very noisy sky estimates in most cases. Alternatively, taking the inverse covariance of the noise, i.e., $\mat{W}=\mat{N}^{-1}$, yields the \textit{maximum likelihood} (ML) mapmaker. This estimator is unbiased and optimal (has the lowest estimation error among all linear unbiased estimators) under Gaussian noise. This often entails a high numerical cost.

More generally, Eq.~\eqref{eqn:template_mapmaking} describes \textit{template} mapmaking in which we model the data to contain extra modes described by the template matrix $\mat{T}$. Although the template amplitudes $\vv{x}$ are unknown, \eqref{eqn:template_mapmaking} allows estimation of $\hat{\vv{s}}$ without explicitly fitting for $\vv{x}$. Mathematically, this is done by \textit{deprojection}; the matrix $\mat{F}_\mat{T}$ in \eqref{eqn:template_mapmaking_FT} projects the data onto the subspace orthogonal to $\mat{T}$'s column space~\footnote{Here, the orthogonality is defined using the weight matrix $\mat{W}$: $\langle \vv{x}, \vv{y}\rangle \equiv \vv{x}^\top \mat{W} \vv{y}$.}. Indeed, $\mat{F}_\mat{T} \mat{T}=0$ and $\mat{F}_\mat{T}\vv{y}=\vv{y}$ for all $\vv{y}$ orthogonal to $T$. It follows that $\mat{F}_\mat{T}^2 =\mat{F}_\mat{T}$, and $\mat{F}_\mat{T}^\top \mat{W} \mat{F}_\mat{T} = \mat{W} \mat{F}_\mat{T}$ and $\mat{F}_\mat{T}^\top \mat{W} = \mat{W} \mat{F}_\mat{T}$. This latter relation ensures that the system matrix in Eq.~\eqref{eqn:template_mapmaking} is symmetric and positive semi-definite given any positive definite $\mat{W}$.

We will refer to $\mat{F}_\mat{T}$ as \textit{template deprojector} throughout this paper. Template mapmaking is equivalent to the maximum likelihood mapmaking on template-deprojected data, if the weight matrix captures the noise covariance correctly. In many applications, the weight matrix is assumed to be diagonal, while the long correlated modes present in the data are modelled as templates. This allows for efficient mitigation of the variance of the estimate at a reasonable numerical cost. This class of template mapmakers is mathematically equivalent to classical destripers, however the sky signal estimate is derived here in a single step without solving for the template amplitudes $\vv{x}$ first.

\subsection{POMME} \label{sec:pomme}

POMME is a template mapmaker with a set of templates chosen to capture all unpolarised signals that vary slowly over time. The template matrix $\mat{T}$ adopted in POMME comprises columns with elements spanning a fixed interval of $\tau$ samples set to $1$, and all other elements vanishing, i.e., 
\begin{align}
    T^{(i)}_t = \left\{
\begin{array}{ll}
      1 & \quad i\tau\le t < (i+1)\tau \\
      0 & \quad \mathrm{otherwise}. \label{eqn:pomme_template}
\end{array} 
\right.
\end{align}
Each detector has its own set of templates, so that the total number of columns in $\mat{T}$ equals $n_\mathrm{detectors} \lfloor n_\mathrm{samples}/\tau \rfloor$, where $\lfloor\cdot\rfloor$ denotes the integer part. Any incomplete intervals at the end of the TOD are masked away. The size of the template amplitude vector $\vv{x}$ is therefore roughly $1/\tau$ times the size of the TOD.

The POMME templates encompass all unpolarised signals independent of their origin, which also includes the unpolarised sky signal.
The latter is not fitted, which avoids possible degeneracies with the template columns, while allowing the value of $\tau$ to be tuned appropriately in order to ensure the best performance of the method as discussed later.
The data model assumed by POMME reads
\begin{align}
    \vv{d} = \mat{P}_{Q/U}\vv{s}_{Q/U} + \mat{T} \vv{x} + \vv{n},   \label{eqn:pomme_data_model}
\end{align}
where the subscripts $Q/U$ indicate to only take the parts for the Stokes $Q$ or $U$ parameters.  The POMME mapmaking equation is given as follows.
\begin{align}
    \hat{\vv{s}}_{Q/U}^\mathrm{POMME} = (\mat{P}_{Q/U}^\top \mat{W} \mat{F}_\mat{T} \mat{P}_{Q/U})^{-1} \mat{P}_{Q/U}^\top \mat{W} \mat{F}_\mat{T} \vv{d}.
    \label{eqn:pomme_mapmaking_equation}
\end{align}

POMME has one tunable parameter $\tau$, which determines the interval length in columns of $\mat{T}$, and hence the typical frequency ranges of $\mat{T}\vv{x}$. A smaller $\tau$ yields more numerous templates with shorter intervals, which can capture a wider range of frequencies and represent them more accurately. Conversely, a larger $\tau$ gives fewer templates and longer intervals, limiting its coverage to lower frequencies and reducing the chance of shrinking the target signal amplitude.

In the extreme case of $\tau=2$, POMME would be equivalent to differencing the measurements in each pair of consecutive samples. This would efficiently remove all contributions that vary much more slowly than the sampling rate ($\mathcal{O}(10^2)\,\mathrm{Hz}$), including the unpolarised signals from sky, atmosphere, and ground. However, the same template deprojection also suppresses the polarised sky signal and leads to large statistical uncertainties in the map estimates. Indeed, a typical HWP rotation frequency of $2\,\mathrm{Hz}$ is substantially less than the sampling rate, and subtracting two consecutive measurements would annihilate the bulk of the polarised signal. On the other hand, the other extreme case of $\tau=n_\mathrm{samp}$, the number of samples, is simply equivalent to binned mapmaking on a TOD with the detector mean subtracted. Consequently, there is no mitigation of the unpolarised signals in this case beyond the effects due to binning the samples in the sky pixels.

The choice of the value of $\tau$ is therefore crucial for the algorithm's performance and is discussed in some detail in Section~\ref{sec:choosing_tau}. Heuristically, however, we can anticipate that the suitable $\tau$-interval length is as short as a typical pixel crossing time (so that the super-pixel modes are removed), but should not be much shorter than the HWP period (so that the polarised signal is preserved). This implicitly imposes some constraints on $\tau$ regarding the experiment and its operations. Denoting the sample rate as $f_\mathrm{samp}$ and the characteristic frequency of POMME as $f_\tau\equiv f_\mathrm{samp}/\tau$, any signals with frequency $f \ll f_\tau$ are well approximated by $\mat{T}\vv{x}$ for some $\vv{x}$.

We assume that, with $\mat{T}\vv{x}$ in the data model \eqref{eqn:template_mapmaking_model} encapsulating all signals that vary slowly over time, the remaining noise component can be well approximated as white and uncorrelated between detectors: $\mat{N}_d=\sigma_d^2 \mat{I}$ for each detector $d$. We therefore take as the weight matrix $\mat{W}_d = \mat{N}_d^{-1}$. With this simplification, the columns of $T$ are orthogonal to each other, and the template deprojector for POMME is simply given by
\begin{align}
    \mat{F}_\mat{T} = \mat{I} - \mat{T}(\mat{T}^\top \mat{N}^{-1} \mat{T})^{-1} \mat{T}^\top \mat{N}^{-1} = \mat{I} - \frac{1}{\tau} \mat{J},
    \label{eqn:pomme_deproj_general}
\end{align}
where $\mat{J}$ is a block-diagonal matrix that consists of $\tau\times\tau$ all-ones matrix blocks $\mat{J}_\tau$. Hence, $\mat{F}_\mat{T}$ subtracts the sample mean evaluated at every $\tau$-interval:
\begin{align}
    (\mat{F}_\mat{T} \vv{d})_t = d_t - \frac{1}{\tau}\sum_{s=i\tau}^{(i+1)\tau-1} d_s,  \label{eqn:pomme_deprojector}
\end{align}
for $i$ satisfying $i\tau \le t < (i+1)\tau$.

We note that, under some special circumstances, POMME reduces down to the usual binned mapmaking estimator without templates: a) each sky pixel is observed by the same number of samples $n_\mathrm{cross}$ at each crossing, b) the HWP rotates an integer amount during $4n_\mathrm{cross}$ samples, c) the HWP rotation frequency remains constant throughout the scan, and d) the weights matrix $\mat{W}$ is diagonal. Under these conditions, the zero-frequency mode in the TOD does not affect polarisation estimates for the binned mapmaker since $\mat{P}^\top_{Q/U} \mat{W} \vv{1}=\vv{0}$. The optimal choice of the POMME templates would then require $\mat{P}_{Q/U}^\top \, \mat{T} \, = \,0$, so that $\mat{P}^\top_{Q/U} \mat{F}_\mat{T} = \mat{P}^\top_{Q/U}$. This is achieved when $\tau=n_\mathrm{cross}/m$ for some integer $m$. POMME is equivalent to a binned mapmaker for such $\tau$: $\hat{\vv{s}}^\mathrm{POMME}_{Q/U} = \hat{\vv{s}}^\mathrm{binned}_{Q/U}$.

In realistic scenarios, the non-uniformity of pixel crossing times means that we cannot rely on the automatic cancellation of the zero- and low-frequency modes as above. Given the huge unpolarised power present at low frequencies in real observations, incomplete mitigation of it can lead to significant leakages into polarisation. POMME aims to remedy the situation by estimating this unpolarised power liable to leak, as an average over the suitably chosen template intervals and subtracting that power from $\mat{P}_{Q/U}^\top \, \vv{d}$ for each crossing.

Consequently, POMME can be looked at as a detrending algorithm, where a local mean is subtracted from each sample. Alternatively, it can be considered as a particular kind of destriper. Unusual, inasmuch as many of the templates contribute only to a single sky pixel and their amplitudes are constrained not owing to their overlap on the sky but to the presence of the continuously rotating HWP. POMME thus relies on the HWP to differentiate between the polarised and unpolarised contributions. It is thus similar in this respect to the so-called `lock-in demodulation' techniques ~\cite{Johnson_2007,Kusaka2014,Ritacco_2016}. However, while these approaches first demodulate the time streams and then bin them into maps without taking into account the effect of the filter, POMME explicitly corrects for any modes deprojected from the data and creates unbiased maps.

Regardless of the adopted perspective, in the POMME algorithm, the unpolarised sky signal (and thus also the total intensity) is always unrecoverably lost during template amplitude marginalisation. However, the polarised sky signal can be efficiently and robustly estimated, as we show in the following Sections. POMME can achieve these goals only thanks to the presence of a rapidly rotating HWP.

\begin{figure}
    \centering
    \includegraphics{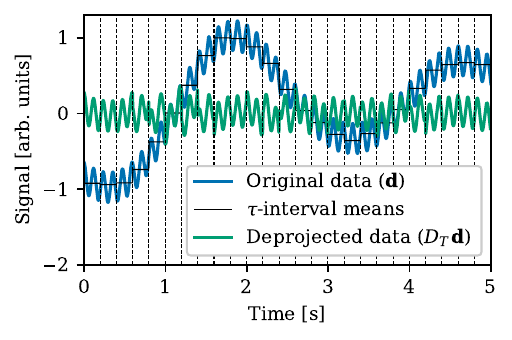}
    \caption{Illustration of the POMME deprojection operator Eq.~\eqref{eqn:pomme_deprojector} applied to synthetic time-ordered data. Mean values within each interval of length $\tau$ are subtracted from the data during the deprojection. The result, shown in green, has low-frequency trends removed while retaining high-frequency oscillations.}
    \label{fig:pomme_deprojection_example}
\end{figure}

As an example, in Figure~\ref{fig:pomme_deprojection_example}, we show the impact of the POMME template deprojector on the time domain data. We generate time-ordered data with $1000$ samples measured in $5$ seconds, and inject two signals: 1) a random Gaussian process with a $0.5$s correlation length and amplitude $1$, and 2) a simple sine wave oscillating at $8$Hz with amplitude $0.2$. POMME deprojection for $\tau=40$ is shown. The $\tau$-interval means $(1/\tau)\mat{J}\vv{d}$ (black solid lines) are subtracted from $\vv{d}$ (blue) to give the deprojected result $\mat{F}_\mat{T}\vv{d}$ (orange). The first signal with characteristic frequency $f\approx 0.3$Hz $\ll f_\tau=5$Hz is heavily suppressed, while the second, $8$Hz-signal survives.

The boxcar templates adopted here lead unavoidably to discontinuities induced by $\mat{F}_\mat{T}$ at each $\tau$-interval boundary. These discontinuities convert some low-frequency power into higher-frequency noise. The aliased noise, if not included in the noise model, can be misconstrued as I-to-P power leakage, and we discuss it in Section~\ref{sec:frequency_response}. However, as Eq.~\eqref{eqn:pomme_mapmaking_equation} demonstrates, the polarised content of the data is unaffected by the templates, at least as long as $\mat{F}_\mat{T}^\top \mat{W} \mat{F}_\mat{T}$ is non-singular, and is fully recovered by the procedure.

We emphasise that the POMME algorithm is not tied to any specific choice of templates. The boxcar templates, Eq.~\eqref{eqn:pomme_template}, have been chosen here for their simplicity and computational efficiency. In principle, any set of templates that accurately captures low-frequency signals can be used. For example, one can additionally introduce terms that are linear in time within each $\tau$-interval.

The system matrix in Eq.~\eqref{eqn:pomme_mapmaking_equation} is expected to be somewhat non-diagonal in the pixel dimension, with off-diagonal couplings arising from templates covering a few neighbouring pixels. It can therefore be efficiently inverted using the preconditioned conjugate gradient method. We note that the system matrix provides the full characterisation of the map noise properties only if the time-domain noise (after deprojection) is white. 

On the algorithmic level, POMME offers a number of potential advantages, including:
\begin{itemize}
    \item \textit{Robustness to unpolarised signals}. All low-frequency contaminant signals are suppressed by template deprojection, while map estimates remain virtually unchanged. POMME can therefore be expected to be robust to systematic contributions from atmospheric emission and unpolarised ground pickup. Similarly, it should remove any unpolarised sky-stationary signals without any assumption that those can be pixelized in sky coordinates, avoiding one potential source of intensity-to-polarisation leakage.
    
    \item \textit{Computational efficiency}. The deprojection operation, Eq.~\eqref{eqn:pomme_deprojector}, scales linearly with the data size, which is significantly faster than, for example, the Toeplitz matrix application of ML mapmakers (typically $\mathcal O(N \log N)$, or $\mathcal O(N \log \lambda)$ for a Toeplitz matrix with bandwidth $\lambda$). Furthermore, the numerical inversion in Eq.~\eqref{eqn:pomme_mapmaking_equation} requires only a small number of iterations because of the small correlations between sky pixels in the estimated maps.
    
    \item \textit{Unbiased estimation}. Despite marginalising over a large number of degrees of freedom, the system matrix $\mat{P}^\top \mat{N}^{-1} \mat{F}_\mat{T} \mat{P}$ remains invertible because all signals of interest live in a different, high-frequency domain. POMME thus delivers unbiased estimates of the polarised sky maps.
\end{itemize}

On the downside, POMME, by its very design, cannot yield an estimate of the unpolarised sky signals.

We also note that owing to its maximum likelihood provenance the POMME algorithm is very flexible and capable of accommodating a number of extensions needed in order to meet the demands of real data. We leave an exploration of these opportunities to future work, and hereafter we focus on the basic formulation and verify the expectations listed above using realistic simulations.

\subsection{Time-domain analysis} \label{sec:frequency_response}

The POMME algorithm is designed to suppress low-frequency contributions due to noise and/or contaminants by deprojecting the proposed templates. While the algorithm deprojects any signal living in the space spanned by the templates, the realistic contaminants are bound to be more complex and are therefore only partially mitigated. In this section, we detail the effects of the deprojection on TODs, containing time-stationary, low-frequency signals, from both analytical and numerical perspectives.

As the POMME deprojection operator is not time-stationary, the noise covariance of the deprojected TOD, defined as,
\begin{align}
\mathcal{N} \equiv \mat{D}_\mat{T} \langle \vv{n} \vv{n}^\top\rangle \mat{D}_\mat{T}^\top
\label{eqn:depNoiseCov}
\end{align}
is not Toeplitz, and its Fourier representation is not diagonal. This is true irrespective of the assumption within this section that the total noise covariance, $\langle \vv{n} \vv{n}^\top \rangle$, is diagonal. In this case, the deprojected noise power spectrum, $\tilde{\mathcal{P}}(f)$, is given by (see Appendix~\ref{appendix:depNoisePS} for the derivation),
\begin{widetext}
\begin{align}
\tilde{\mathcal{P}}&(f_i) \;   = \;  
\mathcal{P}(f_i) \left( 1\, -  \, \genfrac{}{}{}{0}{1}{\tau^2} \, \genfrac{}{}{}{0}{\sin^2 \genfrac{}{}{}{0}{\pi \tau i}{n_{\mathrm{t}}}}{\sin^2 \genfrac{}{}{}{0}{\pi i}{n_{\mathrm{t}}}}\right)^2 \, + \; \genfrac{}{}{}{0}{1}{\tau^4} \, \genfrac{}{}{}{0}{\sin^2 \genfrac{}{}{}{0}{\pi \tau i}{n_{\mathrm{t}}}}{\sin^2 \genfrac{}{}{}{0}{\pi i}{n_{\mathrm{t}}}}\, \sum_{\genfrac{}{}{0pt}{1}{k=-\Big\lfloor\genfrac{}{}{}{1}{n_{\mathrm{t}}/2+i-1}{\lfloor n_{\mathrm{t}}/\tau\rfloor}\Big\rfloor}{k\neq 0}}^{\Big\lfloor\genfrac{}{}{}{1}{n_{\mathrm{t}}/2-i}{\lfloor n_{\mathrm{t}}/\tau\rfloor}\Big\rfloor} \mathcal{P}(f_{|i+k\lfloor n_{\mathrm{t}}/\tau]|})\,\genfrac{}{}{}{0}{\sin^2 \genfrac{}{}{}{0}{\pi \tau (i+k\lfloor n_{\mathrm{t}}/\tau\rfloor)}{n_{\mathrm{t}}}}{\sin^2 \genfrac{}{}{}{0}{\pi (i+k\lfloor n_{\mathrm{t}}/\tau\rfloor)}{n_{\mathrm{t}}}}.
\label{eqn:genDepNoisePS}
\end{align}
\end{widetext}
where $n_\mathrm{t}=n_\mathrm{samp}$ denotes the TOD length, and $f_i \equiv(i/n_\mathrm{t}) f_\mathrm{samp}$ is the $i$th frequency mode in the Fourier series.

The first term on the right-hand side of Eq. \eqref{eqn:genDepNoisePS} expresses the modulation of the power as a function of frequency. This term contributes to the cases of correlated and white noise and reflects the fact that the deprojection suppresses some power and induces correlations locally. For low frequencies $i\ll n_\mathrm{t}$, the leading-order contribution is $\approx (\pi^4/9) (f_i/f_\tau)^4 \mathcal{P}(f_i) $, where $f_\tau\equiv f_\mathrm{samp}/\tau$. We therefore expect low frequencies with $f\ll f_\tau$ to be strongly suppressed by the deprojection.

The second term of Eq. \eqref{eqn:genDepNoisePS} potentially leads to aliasing of the power between different frequency modes. Although this term is suppressed by an additional factor of $\tau^{-2}$ compared to the first term, it can still be significant since there are $\mathcal O(\tau)$ terms in the summation and $\mathcal{P}(f)$ can vary drastically over frequency. The impact of this term is in general heavily dependent on the assumed initial noise spectrum and is more pronounced when more low-frequency power is present. Note that the summation index $k$ of Eq. \eqref{eqn:genDepNoisePS} can be negative, which sometimes causes the leakage from frequency at index $|i+k\lfloor n_\mathrm{t}/\tau \rfloor|=-i+(-k)\lfloor n_\mathrm{t}/\tau \rfloor$.

In the case of the white noise, i.e., $\mathcal{P}(f_i) = \sigma^2 = \mathrm{const.}$, the deprojected noise spectrum reads,
\begin{align}
\tilde{\mathcal{P}}_{\mathrm{white}} \;   = \;  \sigma^2 \left( 1\, -  \, \genfrac{}{}{}{0}{1}{\tau^2} \, \genfrac{}{}{}{0}{\sin^2 \genfrac{}{}{}{0}{\pi \tau i}{n_{\mathrm{t}}}}{\sin^2 \genfrac{}{}{}{0}{\pi i}{n_{\mathrm{t}}}}\right).
\label{eqn:whiteDepNoisePS}
\end{align}

\begin{figure}
    \begin{center}
    \includegraphics [width=1.0\columnwidth]
    {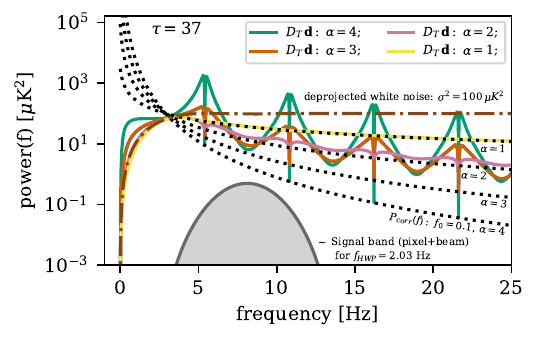}
    \includegraphics[width=1.0\columnwidth]
    {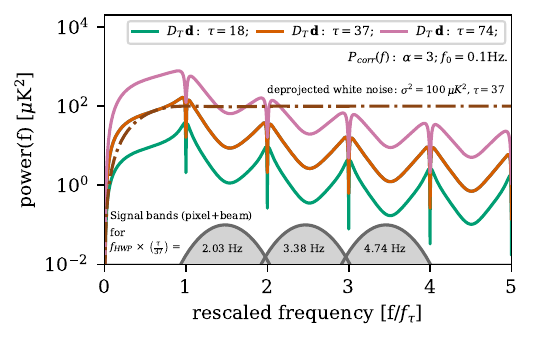}\\
    \phantom{x}\includegraphics[width=1.0\columnwidth]
    {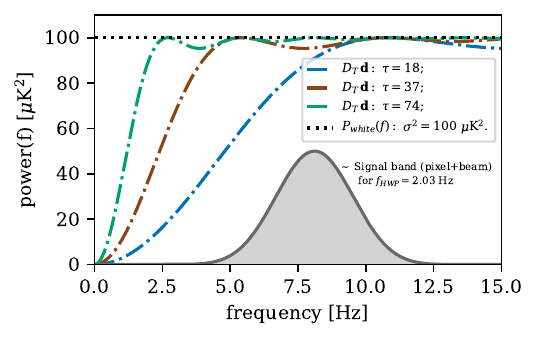}
    \end{center}
    \vskip -0.5truecm
    \caption{
    The power spectra of the deprojected noise, $\mat{D}_\mat{T}\,\vv{n}$, for different input noise models. The solid lines show the correlated noise cases with parameters as defined in each panel. The dotted lines display the corresponding input spectra. The dot-dashed lines depict the deprojected white noise spectra for the input white noise level of $100\mu$K$^2$. Note that the dot-dashed dark brown line corresponds to the same case in all three panels, providing a reference. The relative normalisation of the correlated and white noise spectra is such that the coadded input noise would have $f_{\mathrm{knee}} = 3$Hz. The shaded areas depict idealised combined beam and pixel window function corresponding to the value of $f_{\mathrm{HWP}}$ as indicated in each panel. The top panel shows the results for different correlated noise parameters for a fixed value of $\tau =37$. The middle panel shows the results for the same input noise spectrum and $\tau$ set to multiples of $18$.
    The bottom panel shows results for white noise and different values of $\tau$.
    }
    \label{fig:pomme_filtered_PSD}
\end{figure}
This formula combines the contributions from the modulation and the aliasing, as given by the first and second terms of Eq.~\eqref{eqn:genDepNoisePS}, respectively.
We show the resulting spectra in Fig.~\ref{fig:pomme_filtered_PSD} and discuss them in detail below.

For concreteness, we consider the correlated noise described by a power law given by,
\begin{align}
\mathcal{P}_{\mathrm{corr}} \; \propto \; \left(
\genfrac{}{}{}{0}{f_0}{f+f_0}\right)^\alpha,
\end{align}
where $f_0$ defines the scale at which the noise power flattens at the low frequency end. The white noise is defined as before by the frequency-independent noise variance, $\sigma^2$. The top panel of Fig~\ref{fig:pomme_filtered_PSD} shows the input correlated noise spectra (dotted lines) and the corresponding deprojected noise spectra (solid lines) for values of $\alpha$ ranging from $1$ up to $4$. The deprojected noise spectrum corresponding to the white noise with $\sigma^2 = 100\mu K^2$ is also shown. The relative normalisation of the correlated and white noise spectra is such that the effective knee frequency of the coadded input noise would be set to $f_{\mathrm{knee}} = 3$Hz, a value typically associated with the atmospheric emissions.

We can make the following observations:
\begin{itemize}
     \item \textit{Suppression of low-frequency signal}. As expected, the low-frequency modes in the TOD are reduced by a factor $\propto (f/f_\tau)^4$ at the PSD level, where $f_\tau \equiv f_\mathrm{samp}/\tau$. This is important as these modes are the low-frequency noise that leaks into the polarisation estimates of a binned mapmaker if left unmitigated, as discussed in Section~\ref{sec:pomme}.
     Furthermore, the noise power at zero frequency in the deprojected time stream always vanishes, i.e., $(\mathcal{F} \,\mathcal{N}\,\mathcal{F}^\dagger)_{00}=0$, as expected given the POMME deprojection operator removes the constant mode. Similarly, the power initially present at zero frequency, i.e., $\mathcal{P}_{\mathrm{corr/white}}(0)$, is always removed and does not ever contribute to the power of the deprojected noise.
     
     \item \textit{Transfer of some low-frequency power to harmonic modes.} Due to discontinuities at $\tau$-interval boundaries which are present in the POMME templates but absent in the actual low-frequency signal, the deprojection inevitably transfers power between modes that are spaced by multiples of $f_\tau$ apart. This effect is particularly pronounced if sufficient low-frequency power excess, $\alpha \ge 1$, is present and results in aliasing of some of that power to the modes in the vicinity of the harmonics $mf_\tau$, $m=1,2,\cdots$. However, we note that precisely at the harmonics, the power of the deprojected and input time streams is exactly the same and equal to $\mathcal{P}_{\mathrm{corr}}(mf_\tau)$, giving rise to the conspicuous dips shown in the deprojected spectra in Fig.~\ref{fig:pomme_filtered_PSD}. The amount of the aliased power strongly depends on $\alpha$, but also on the assumed spectrum flattening scale, $f_0$. In the extreme cases, it can exceed the white noise floor by orders of magnitude, however it remains very localised in the immediate vicinity of the harmonics of $f_\tau$, and it is usually significantly below the white noise level in between. 
     
     The aliased power also depends on the value of $\tau$, as shown in the middle panel of Figure~\ref{fig:pomme_filtered_PSD}. The horizontal axis is rescaled so the corresponding harmonics appear in the same position, independent of the value of $\tau$. While the qualitative behaviour is similar for all values of $\tau$, it is clear that overall the power for any fixed value of $f/f_\tau$ increases with the value of $\tau$. This is intuitively understandable as the templates with larger $\tau$ can only capture progressively longer modes, and more residual low-frequency power remains in the deprojected time stream.
     
     \item \textit{Modulation of the white noise weights at intermediate and high frequencies.} The third panel of Figure~\ref{fig:pomme_filtered_PSD} shows the deprojected white noise cases for different values of $\tau$. While the low frequencies are significantly suppressed as discussed above, the power at frequencies higher than $f \sim 1/4 f_{\mathrm{HWP}}$ remains essentially constant with small, $\sim 2-5\%$, modulations imprinted on it.
\end{itemize}

\begin{figure}
\begin{center}
    \includegraphics[width=0.975\columnwidth]
    {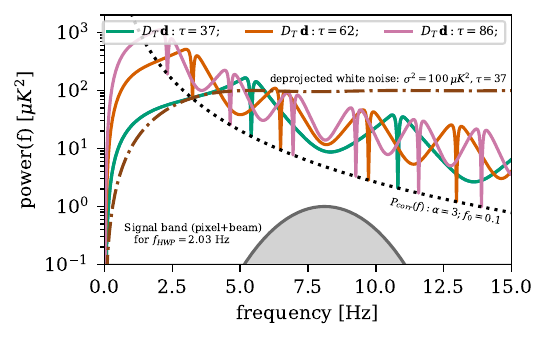}\\
    \phantom{i}\includegraphics[width=0.975\columnwidth]
    {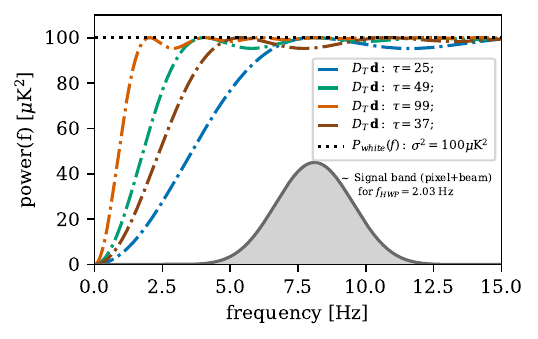}\\
    \phantom{xi}\includegraphics[width=0.9\columnwidth]
    {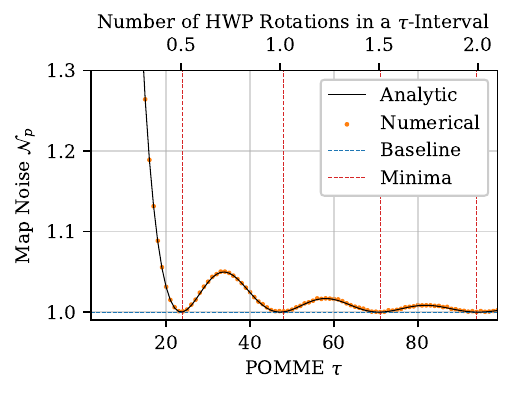}
\end{center}
\vskip -0.75truecm
    \caption{A pictorial summary of the $\tau$ selection heuristics proposed in this work, Section~\ref{sec:choosing_tau}, for an experiment with continuously rotating HWP at $f_{\mathrm{HWP}} = 2.03$Hz and the expected signal band as approximately depicted by the pixel/beam kernel (shaded areas). The top panel shows the spectra of the correlated residuals after the deprojection (solid lines), with values of $\tau$ satisfying Eq.~\eqref{eqn:condLeak} minimising the correlated residual. The input noise (dotted line) has $\alpha = 3$ and $f_0=0.1$Hz. The middle panel displays the deprojected white noise weights for values of $\tau$ given by Eq.~\eqref{eqn:condWhite} maximising the weights in the signal band (dot-dashed lines), while minimising the aliasing. The dark brown line shows the deprojected white noise weights for a value of $\tau$ minimising the correlated residual, as also shown in the top panel. The bottom panel shows the pixel noise as a function of $\tau$ computed analytically, Eq.~\eqref{eqn:radek_magic_formula} (black line) and via simulations (orange dots). The vertical axis is normalised to the best achievable noise level, where no modes are deprojected. Clearly, the POMME algorithm can recover the best possible sky map if $\tau$ is set as in Eq.~\eqref{eqn:condWhite} (middle panel), but the precision loss is minor if instead the condition in Eq.~\eqref{eqn:condLeak} is satisfied (top panel).}
    \label{fig:pomme_information_loss}
\end{figure}

\subsection{Choosing $\tau$} \label{sec:choosing_tau}

The choice of $\tau$ is critical to POMME's performance. As we hinted at in the previous Section, $\tau$ determines both the sensitivity of the produced polarisation maps as well as the level of low-frequency residual they contain. We discuss both these aspects in some detail in this Section and provide quick, easy heuristics, which we then validate through end-to-end runs in the follow-up Section.

Generally, we would like to choose $\tau$ that, for a given experimental setup, minimises: 1) the map-level residuals due to the correlated noise, and 2) the map-level noise due to the uncorrelated noise. The former requirement can be satisfied if the time-domain correlated residuals are low in the frequency bands where the sky signal resides. While, in general, the sky response in the frequency domain can be very complex, for experiments featuring the HWP we expect the polarised signal to reside in frequency bands that roughly correspond to the scan harmonics, shifted to be centred around $4f_{\mathrm{HWP}}$. These are furthermore modulated by the pixel/beam kernel reflecting the sky signal smoothing on the beam and pixel scales as schematically shown in Fig.~\ref{fig:pomme_filtered_PSD}. The breadth of the kernel will depend on some specific details, like pixel crossing time, relative size of the pixel and the beam, etc. Many of which are fixed for any particular experiment and are beyond the data analyst's control. However, while these may affect the performance of the method, they will not affect the best choice of $\tau$ in general.

Adopting this perspective, we can infer from  Fig.~\ref{fig:pomme_filtered_PSD} that a suitable value of $\tau$ should be such that the signal band peaks between two consecutive harmonics of $f_{\tau}$. As the band is centred around $4\,f_{\mathrm{HWP}}$, we want that value to be roughly halfway between multiples of $f_{\tau}$. This can be realised when 
\begin{align}
    \tau_{m+1/2} \equiv \left( m + \frac{1}{2} \right) \Big\lfloor \frac{f_\mathrm{samp}}{4f_\mathrm{HWP}}
    \Big\rfloor, \quad m=1,2,\dots,
    \label{eqn:condLeak}
\end{align}
as is shown in the top panel of Fig.~\ref{fig:pomme_information_loss}. We see that choosing the lowest value of $\tau$ corresponding to $m=1$ seems to be optimal, ensuring the lowest level of the residual. This is consistent with our discussion of Fig.~\ref{fig:pomme_filtered_PSD} in the previous Section.

For the map noise level, the heuristics are more involved here as we need to consider not only the deprojected white noise, but also how the POMME deprojector affects the sky signal. In fact, within the signal frequency band, as depicted by the beam+pixel kernels in Figs.~\ref{fig:pomme_filtered_PSD} and~\ref{fig:pomme_information_loss}, both the sky signal and the white noise are always impacted similarly, regardless of the value of $\tau$. The white noise is, however, present across the full range of frequencies, unlike the sky signal, which is confined to the signal frequency band. The values of $\tau$ which can minimise the aliasing will also minimise the noise relative to the sky signal and should thus be preferred as far as the map noise is concerned.

The amount of aliasing is quantified by the second term in Eq.~\eqref{eqn:genDepNoisePS} and can be easily computed as a difference of the deprojected white noise spectrum in Eq.~\eqref{eqn:whiteDepNoisePS} and the first term on the right-hand side in Eq.~\eqref{eqn:genDepNoisePS}, giving,
\begin{align}
    \genfrac{}{}{}{0}{1}{\tau^2} \, \genfrac{}{}{}{0}{\sin^2 \genfrac{}{}{}{0}{\pi \tau i}{n_{\mathrm{t}}}}{\sin^2 \genfrac{}{}{}{0}{\pi i}{n_{\mathrm{t}}}}
    \,
    \left(1\, - \,
    \genfrac{}{}{}{0}{1}{\tau^2} \, \genfrac{}{}{}{0}{\sin^2 \genfrac{}{}{}{0}{\pi \tau i}{n_{\mathrm{t}}}}{\sin^2 \genfrac{}{}{}{0}{\pi i}{n_{\mathrm{t}}}}
    \right).
\end{align}
For the centre of the signal band $i/n_t \simeq 4\, f_{\mathrm{HWP}}/f_{\mathrm{samp}}$, and this is minimised when
\begin{align}
\tau_m \; = \; m\,\Big\lfloor \genfrac{}{}{}{0}{f_\mathrm{samp}}{4\,f_\mathrm{HWP}}\Big\rfloor, \quad m=1,2,\dots
\label{eqn:condWhite}
\end{align}
We note that, somewhat counterintuitively, this corresponds to the values of $\tau$ which maximise the deprojected white noise power in the signal band.

We can make this connection more directly, albeit under some simplifying assumptions. Let us assume that each $\tau$-interval always corresponds to a single pixel on the sky, and thus there are no templates crossing over between two different pixels. If so, the pixel-pixel noise covariance matrix, given by $ \mathrm{Cov}\left[ \hat{\vv{s}}^\mathrm{POMME} \right] = (\mat{P}^\top \mat{N}^{-1} \mat{F}_\mat{T} \mat{P})^{-1}$, is diagonal if the time domain noise is white, and the diagonal elements are given by,
\begin{align}
\mathcal{N}_p = \frac{2\sigma^2}{n_\mathrm{hits}} \left( 1 - \frac{1}{\tau^2}\frac{\sin^2 \tau 
\genfrac{}{}{}{0}{4\pi f_\mathrm{HWP}}{ f_\mathrm{samp}}
}{\sin^2\genfrac{}{}{}{0}{4\pi f_\mathrm{HWP}}{ f_\mathrm{samp}}} \right)^{-1},
\label{eqn:radek_magic_formula}
\end{align}
where $n_{\mathrm{hits}}$ denotes the number of hits per pixel.
We see therefore that whenever the condition in Eq.~\eqref{eqn:condWhite} is fulfilled, the pixel domain noise is as good as the case without any template deprojection. POMME is therefore lossless in these cases. This is not really surprising, as in these cases $\mat{P}^\top_{Q/U}\,\mat{T} = 0$, and the templates do not play any role (see also the discussion in Sect.~\ref{sec:pomme}). We plot Eq.~\eqref{eqn:radek_magic_formula} in the last panel of Fig.~\ref{fig:pomme_information_loss}. We see that not only is the pixel domain noise the best possible for the $\tau$ values that satisfy Eq.~\eqref{eqn:condWhite}, but also that the noise gets worse by no more than $\sim 5$\% of the best value for any value of $\tau \ge \tau_1$. On the other hand, for $\tau\ll f_\mathrm{samp}/4f_\mathrm{HWP}$, the map noise scales $\propto \tau/(\tau^2-1)\approx 1/\tau$, which can grow rapidly for smaller values of $\tau$.

At face value, both conditions, as exemplified by Eqs.~\eqref{eqn:condLeak} and~\eqref{eqn:condWhite}, are exclusive and cannot be satisfied simultaneously. However, as noted in the previous Section, the power modulation of the white noise weights at frequencies higher than $1/4f_\mathrm{HWP}$ is merely on the order of at most a few per cent, corresponding to at most $\sim 5$\% losses in the pixel domain noise, Eq.~\eqref{eqn:radek_magic_formula}. Consequently, the gain in sensitivity from optimising the value of $\tau$ is bound to be rather limited. We therefore recast the last requirement as merely $\tau > \tau_1$, so we avoid catastrophic sensitivity loss at the low-frequency part. This is automatically fulfilled if we choose the value of $\tau$ as
\begin{align}
    \tau_{\mathrm{opt}} \simeq \tau_{3/2} \, = \, 
    \,\Big\lfloor \genfrac{}{}{}{0}{3\,f_\mathrm{samp}}{8\,f_\mathrm{HWP}}\Big\rfloor.
    \label{eqn:tauOptimFinal}
\end{align}
We note that while larger values of $m$ could also be adopted to define $\tau$, it seems prudent to keep that value as small as possible in order to avoid excessive pixel-pixel correlations. Similarly, lower values of $\tau$, for instance,  corresponding to $m=0$ in Eq.~\eqref{eqn:condLeak}, could also be of interest as it could ensure superior control of the correlated noise residual. Such low values of $\tau$ tend to oversuppress the signal power, leading to unnecessarily noisy maps. Whether such a trade-off is acceptable may need to be carefully assessed on a case-by-case basis. Overall, while the heuristics presented here are designed to provide a convenient starting point, a more in-depth, simulation-based assessment and/or validation may be warranted in actual applications, as in the analysis presented in Section~\ref{sec:results}.

Eq.~\eqref{eqn:tauOptimFinal} defines our final recommendation concerning the value of $\tau$.

We emphasise here that for any given experiment, a single parameter, however optimised, may not always be sufficient to ensure fully satisfactory performance of the method in all the relevant aspects. Indeed, success in this respect will in general depend on some of the hardware characteristics of the instrument and/or its operations.  Conversely, the considerations presented in this Section can be used as guidance for careful optimisation of the HWP rotation speed, sampling rate, scanning speed, and (beam-size) pixel crossing time that could allow for the method to reach its full potential, thus successfully mitigating the impact of some of the most insidious systematic effects as discussed here.

\section{POMME implementation}
\label{sec:implementation}

\subsection{Codebase}

The POMME mapmaking pipeline is developed in the context of
\furax~\cite{chanial2026}, an open-source Python library designed to construct and manipulate linear operators for solving inverse problems. \furax\ is powered by JAX~\cite{jax2018github}, a public Python library with optimised performance via just-in-time (JIT) compilation, auto-differentiation, and GPU acceleration. \furax\ fully utilises these benefits to solve the mapmaking equations, such as Eqs.~\eqref{eqn:template_mapmaking} and \eqref{eqn:pomme_mapmaking_equation}, efficiently.

The \furax\ package is ideal for POMME mapmaking for several reasons. First, POMME involves many repeated operations with the same input data shape, such as deprojecting TODs. JIT-compilation of these operations can significantly speed up their computation at a small, one-off overhead cost. Second, the modularity of \furax\ allows easy construction and efficient testing of the POMME mapmaker. Third, \furax\ provides observational data interfaces for both \toast\ and \sotodlib, ideal for validation and testing as well as application to the incoming Simons Observatory data. Lastly, the \furax framework is very flexible and can be straightforwardly extended whenever needed, while directly inheriting much of its computational efficiency.

Most operations necessary for mapmaking are already provided in \furax, including,
\begin{itemize}
    \item \textit{On-the-fly pointing}. The full time-dependent pointing information of each detector can be decomposed into a time-varying boresight pointing direction and a fixed per-detector offset. These two pieces of information are combined ``on-the-fly'' to ease the memory requirements, allowing more observations to be mapped simultaneously.
    \item \textit{Preconditioned conjugate gradient}. \furax\ is interfaced with the \texttt{lineax} package, which provides various matrix-free linear solvers including a Preconditioned Conjugate Gradient (PCG) algorithm. Details of the preconditioner are given in Section~\ref{sec:precond}.
    \item \textit{Noise fits}. We fit noise models to the power spectral density (PSD) estimated from data. Currently \furax\ supports the white noise model $P(f)=\sigma_\mathrm{white}^2$, and the $1/f$ noise model ($P(f)=\sigma_\mathrm{white}^2[1 + (f+f_0)/f_\mathrm{knee})^\alpha]$. The optimisation is performed using the \cadre package\footnote{\url{https://github.com/CMBSciPol/CADRE}}, which fully benefits from JAX's autodifferentiation for efficient optimisation.
\end{itemize}

\subsection{Gap treatment}

It is common for a subset of samples within a given observation to be invalid. Examples include intervals during which the telescope changes direction (``turnaround''), passing of planets and point sources, and readout glitches. All invalid samples are flagged during pre-processing steps, leading to gaps in the processed data, which need to be excluded from the calculations during the mapmaking procedure.

Even though POMME deprojection acts effectively as a low-pass filter, the templates are localised in the sample space and are disjoint. This makes the gap treatment especially straightforward: given a TOD-sized binary mask $\vv{m}$, we can define a masking operator $\mat{M}$ such that $(\mat{M}\vv{x})_i = m_ix_i$; the masked samples are simply set to zero. The full mapmaking operation then becomes:
\begin{align}
    \hat{\vv{s}}^\mathrm{POMME} = (\mat{P}^\top \mat{N}^{-1} \mat{D}_\mat{T} \mat{M} \mat{P})^{-1} \mat{P}^\top \mat{N}^{-1} \mat{D}_\mat{T} \mat{M} \vv{d}. \label{eqn:pomme_mapmaking_equation_with_mask}
\end{align}
Note that $\mat{N}^{-1}\mat{D}_\mat{T}\mat{M}$ = $\mat{M}\mat{N}^{-1}\mat{D}_\mat{T}\mat{M}$ as we assume diagonal noise covariance $\mat{N}$. This masking operation has a negligible cost compared to other mapmaking operations.

We make two small adjustments to the sample mask $m$ for POMME. First, any $\tau$-intervals that contain any invalid samples are fully masked away. This removes up to $(\tau-1)\cdot$(number of masked intervals) additional samples, typically less than $0.1\%$ of the total number of samples. Second, any incomplete $\tau$-intervals at the end of the TOD are discarded. This is less than 1 second in a 1-hour observation.

\subsection{Preconditioning \& Pixel selection}

\label{sec:precond}

The system matrix, defined as
\begin{align}
    \mat{S} \equiv \mat{P}^\top \mat{N}^{-1} \mat{D}_\mat{T} \mat{M} \mat{P},
\end{align}
has to be inverted for the map estimation, Eq.~\eqref{eqn:pomme_mapmaking_equation_with_mask}. This matrix, shaped $n_\mathrm{stokes}n_\mathrm{pixels} \times n_\mathrm{stokes}n_\mathrm{pixels}$, has in general non-zero off-diagonal elements; whenever a $\tau$ interval spans across two sky pixels pointed, the corresponding pixels get correlated during deprojection.

A direct inversion of $\mat{S}$ is often computationally infeasible. Therefore, we employ the Preconditioned Conjugate Gradient (PCG) method to solve Eq.~\eqref{eqn:pomme_mapmaking_equation_with_mask} numerically. Note that $\mat{S}=(\mat{D}_\mat{T} \mat{M} \mat{P})^\top \mat{N}^{-1} (\mat{D}_\mat{T} \mat{M} \mat{P})$ is symmetric and positive semi-definite, as required for PCG.

We first compute an approximate system matrix \textit{without} the deprojection step:
\begin{align}
    \mat{S}_\mathrm{approx} \equiv \mat{P}^\top \mat{N}^{-1} \mat{M} \mat{P},
    \label{eqn:approx-system-matrix}
\end{align}
which is a block-diagonal matrix consisting of $n_\mathrm{stokes}\times n_\mathrm{stokes}$ blocks per pixel. This matrix can therefore be inverted block-by-block quickly.

The matrix $\mat{S}_\mathrm{approx}$ is used for two purposes. First, we select sky pixels with good coverage during the observation. The threshold is given by 1) a minimum number of samples corresponding to the pixel, and 2) a maximum condition number of the $\mat{S}_\mathrm{approx}$ block. A sky pixel that has not been observed long enough or with few polarisation angles will be dropped.

Second, the matrix inverse $\mat{S}_\mathrm{approx}^{-1}$ after the pixel selection is used as a preconditioner for the PCG required in Eq.~\eqref{eqn:pomme_mapmaking_equation_with_mask}. The matrix $\mat{S} \mat{S}_\mathrm{approx}^{-1}$ is better conditioned than $\mat{S}$, and the PCG typically converges within 15 iterations at a relative tolerance of $10^{-6}$.

\subsection{Multi-observation mapmaking}

Information from multiple observations can be incorporated into a single sky map by either co-adding multiple sky maps from a subset of observations or through a single mapmaking operation on the full data.
If the data consists of $n_\mathrm{obs}$ observations with independent noise properties, then the POMME equation can be written as
\begin{align}
     \left[ \sum_{i} {\mat{P}^{(i)}}^\top {\mat{N}^{(i)}}^{-1} \mat{D}_\mat{T}^{(i)} \mat{M}^{(i)} \mat{P}^{(i)}\right] \hat{\vv{s}}^\mathrm{POMME} \nonumber \\=  \sum_i {\mat{P}^{(i)}}^\top {\mat{N}^{(i)}}^{-1} \mat{D}_\mat{T}^{(i)} \mat{M}^{(i)} \vv{d}^{(i)} , \label{eqn:pomme_mapmaking_equation_multi}
\end{align}
where the superscript $(i)$ indicates the $i$th observation.

Multi-observation mapmaking in \furax\ requires just two I/O passes over the data stored on disk. The first builds the various operators ($\mat{P}^{(i)}$, $\mat{N}^{(i)}$, $\mat{M}^{(i)}$) of Eq.~\eqref{eqn:pomme_mapmaking_equation_multi} with consistent shapes (in order to re-use compute kernels); the second accumulates the right-hand side of the equation, a pixel-domain object, such that the full TOD vector is never stored in memory for more than one observation at a time. Only minimal information from them, such as the pointing (boresight and detector) and sample mask, is stored within the operators across observations for the full PCG inversion in the end.

Mapping multiple observations in a single operation on the full dataset is critical for a method like POMME. It increases the sky coverage, resulting in more pixels passing the selection criteria detailed in the previous section, and improving the stability of the numerical inverse. This is also beneficial for computational efficiency, given that the PCG convergence rate directly depends on the condition number of the system matrix.

\subsection{Implicit pair differencing} \label{sec:implicit_pair_diff}

We assume that the telescope's focal plane consists of co-located detector pairs, which are sensitive to orthogonal polarisation directions. In general, fitting the noise model directly from the TOD leads to unavoidably somewhat different noise weights for these detector pairs. In some circumstances it may be however well-motivated and beneficial to assume the same noise weights for both detectors in each pair.

Indeed, in the case of perfect, or nearly perfect, instrument calibration, it can be inferred from Eq.~\eqref{eqn:data_model} that unpolarised signals contribute in the same way to the data of each detector in a pair, whereas polarised signals have opposite signs.
Therefore, an application of the transposed pointing operator, $\mat{P}^\top$, to the data stream, $\vv{d}$, on the right-hand side of the mapmaking equation, would naturally lead to (nearly) perfect cancellation of any unpolarised contributions in the polarised sky estimates. This also applies if the weights are allowed for as long as they are taken to be the same for two detectors in each pair. Following~\cite{biquard2025}, we refer to this effect as {\em implicit pair differencing}. While this is clearly a desirable feature of such experimental design and should be capitalised on in applications to actual data, this can be misleading in the context of methodological studies, as the one undertaken in this work, giving unrealistically good estimates unrelated to the new algorithmic extensions proposed here. 

Therefore, to prevent such implicit pair differencing, we manually inject random scatter at a $5\%$ level to the detector weights as used by the POMME algorithm. As discussed in~\cite{biquard2025}, this is indeed large enough to result in substantial atmospheric signal residual in the polarised map estimates if no additional mitigation measures are included.

\subsection{Simulation and analysis setup} \label{sec:simulation_setup}

We use the publicly available \toast\ library\footnote{\url{https://github.com/hpc4cmb/toast}} (Time Ordered Astrophysics Scalable Tools) for simulating ground-based CMB observations, detector noise, and atmospheric emissions. We generated 581 TODs from a configuration analogous to the Simons Observatory Small Aperture Telescope (SAT) using \texttt{toast\_so\_sim} provided within the \sotodlib library\footnote{\url{https://github.com/simonsobs/sotodlib}}. The simulations consist of 1-hour constant-elevation scans with a sampling rate of $200$ Hz, with the HWP rotating at a constant speed of $\approx 2.0024$Hz. This random offset from $2$ Hz prevents any potential alignment with the sampling frequency. To save computational time, we focus our attention on scans targeting $(\mathrm{RA},\mathrm{Dec})=(0,-20)$, and only include 108 detectors per wafer in the focal plane (each TOD consists of one of 7 wafers). 

For simulating the CMB signal, we use the \textit{Planck} SMICA maps at $n_\mathrm{side}=2048$, with the beam window adjusted to 27.4 arcmin to match the SAT specifications. The instrumental noise has been generated using \toast\ for SAT, with a 5\% random scatter per detector to prevent implicit pair differencing as outlined in Section \ref{sec:implicit_pair_diff}.

The atmospheric emission is also simulated with \toast, using both coarse-grid and dense simulations to accurately emulate $1/f$-noise in the TOD arising from the atmosphere. When adding to the observation TOD, we scaled these by a factor of 10, increasing the effective number of observations (or decreasing the average PWV level).

We split the 581 TODs into 8 groups of $\sim72$ observations and perform multi-observation reconstruction of each group independently. The resulting 8 maps are co-added using inverse variance weighting. We restrict our attention to the map pixels that have at least $10^{-2}$ times the maximum number of hits. The final co-added maps cover $\approx5.8\%$ of the sky. HEALPix pixellisation \cite{healpix} with $N_\mathrm{side}=512$ is used for all mapmaking purposes.

For analysis in the harmonic domain, we deploy \namaster\ \cite{alonso2023namaster} to compute pseudo-$C_\ell$ estimates from partial sky. Sky masks apodisation scale is set to 1 degree, and the $\ell$ bin size to 10.

\section{Results}
\label{sec:results}

\subsection{Code validation}

We have thoroughly validated our pipeline.  First, we tested the integrity of our Python implementation in \furax\ through extensive unit tests on various aspects of the code,  including pointing, data interfacing, and linear algebraic operations.

Next, we validated our mapmaking code on noiseless simulations with input CMB only. The input CMB is recovered accurately, with the fractional error $<10^{-3}$ in pixel space, or $O(10^{-6})$ at the power spectra ($C_\ell$) level.

Lastly, we checked our binned mapmaker written in \furax\ against \sotodlib, and found the results to be consistent.

\subsection{Atmosphere mitigation}

Having validated the pipeline, we test the performance of POMME on our simulation suite involving CMB, instrumental noise, and unpolarised atmospheric emission, as detailed in Section \ref{sec:simulation_setup}. At the TOD level, the atmospheric signal completely dominates the CMB by 3-4 orders of magnitude. However, as this unpolarised signal is slowly varying with frequency $\ll 4f_\mathrm{HWP}=8\,\mathrm{Hz}$, we expect POMME templates to capture them effectively.

\begin{figure}[h]
    \centering
    \includegraphics[width=0.5\textwidth]
    {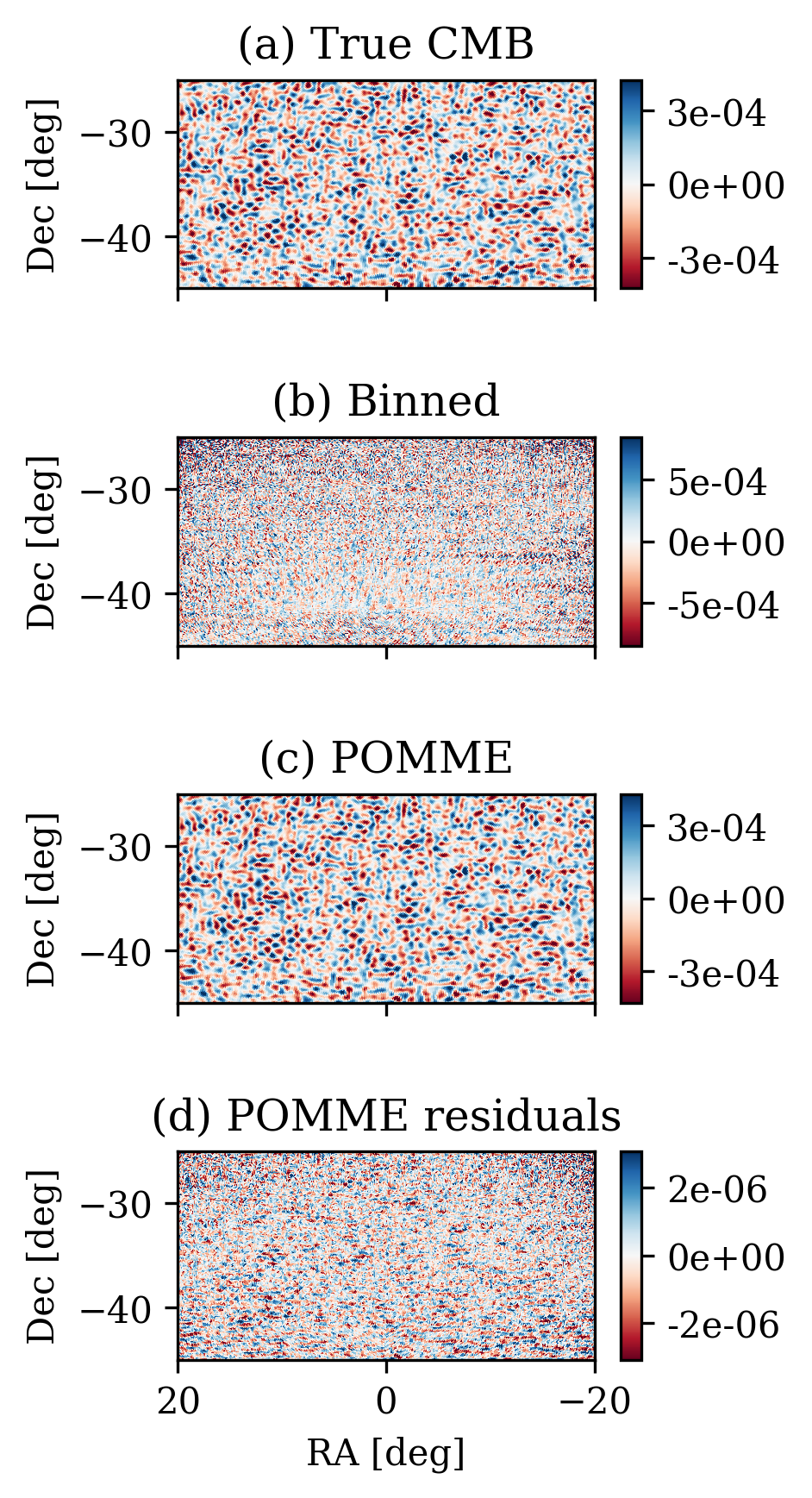}
    \caption{Map estimates obtained from TOD simulations with instrumental noise and atmospheric emission included. The underlying CMB Q map (a) is buried under the large noise when using a conventional binned mapmaker without filtering (b), but POMME with $\tau=37$ (c) effectively mitigates this noise and recovers large-scale polarisation signals. The residual map (d) is mostly white noise. The signal unit is thermodynamic $K$.}
    \label{fig:atm_sim_Q_maps}
\end{figure}

Figure \ref{fig:atm_sim_Q_maps} illustrates the efficacy of POMME in mitigating atmospheric emissions. The CMB Q map signal of $\mathcal O(10^{-6}) \mathrm{K}$ is completely buried in noise for the binned mapmaker. Due to the large amplitude of these contaminating signals, even a small leakage of $\mathcal O(10^{-3})$ can greatly influence the overall noise level of polarisation map estimates. Note the thin stripes in Figure \ref{fig:atm_sim_Q_maps} (b), especially evident at small declinations, which are parallel to the telescope's scan direction. POMME, on the other hand, can accurately recover the large-scale polarisation signal, as shown in Figure \ref{fig:atm_sim_Q_maps} (c). Figure \ref{fig:atm_sim_Q_maps} (d) displays the difference between the estimated map and the underlying map, which is mostly white and does not show correlations along the scanning directions.

\begin{figure*}
    \centering
    \includegraphics
    {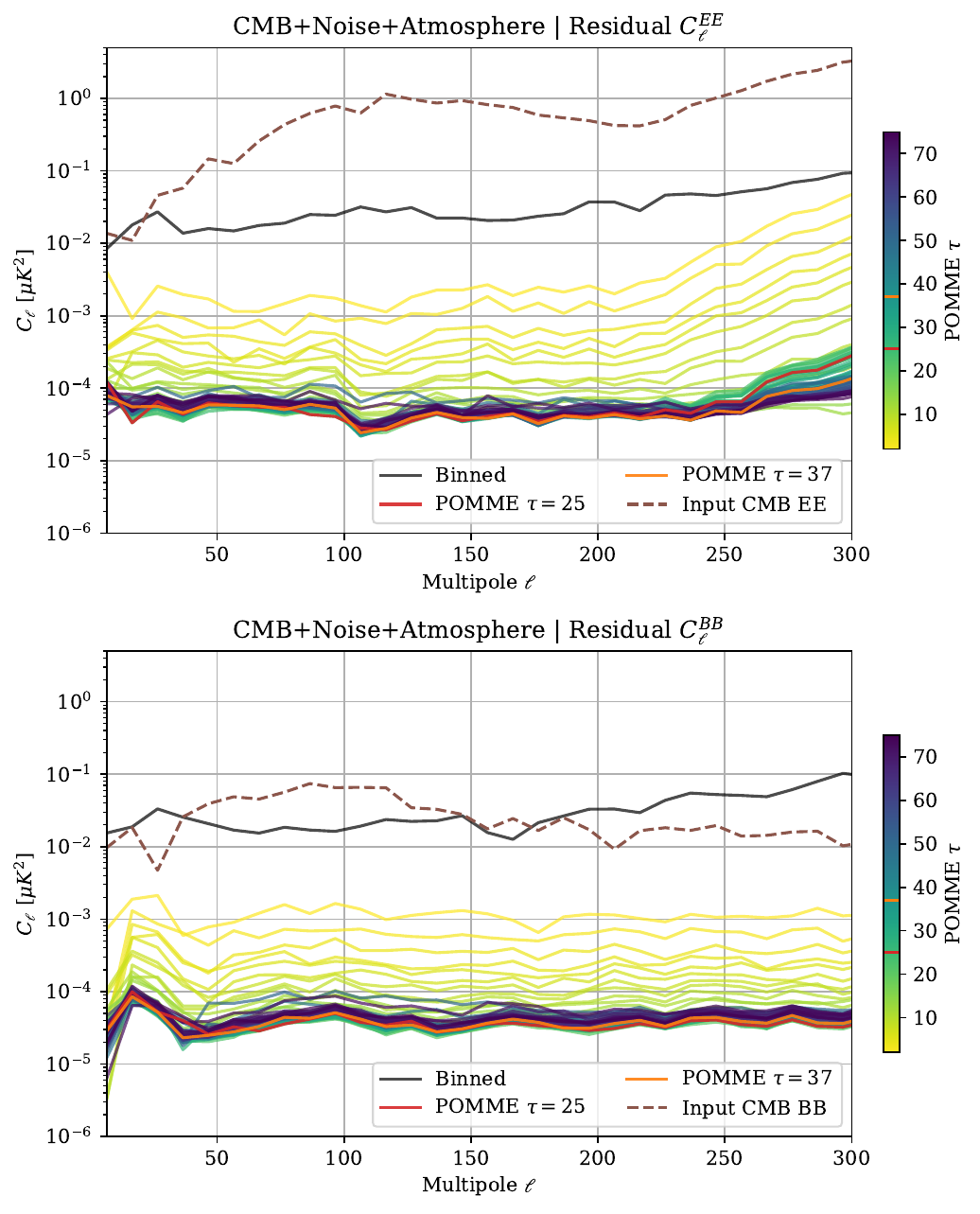}
    \caption{Noise levels on the polarisation power spectrum evaluated from atmospheric simulations. The pseudo-$C_\ell$s of the map estimate residuals against the noiseless CMB map are shown. While the binned mapmaker (black) is severely affected by the atmospheric noise, POMME (coloured) effectively suppresses its effect. The red and orange curves correspond to POMME with $\tau=[ m(f_\mathrm{samp}/4f_\mathrm{HWP})]$ for $m=1$ and $m=3/2$, respectively. }
    \label{fig:atm_auto_minus_cross_tau}
\end{figure*}

The effectiveness of POMME is also evident in harmonic space. Figure \ref{fig:atm_auto_minus_cross_tau} shows the noise levels in the angular power spectrum obtained from the binned mapmaker and POMME. Here, the noise spectra have been computed by differencing the map's auto-spectrum with its cross-spectrum against the underlying CMB map. All qualitative features of Fig \ref{fig:atm_auto_minus_cross_tau} are preserved when studying residual $C_\ell$'s instead.

The simple binned mapmaker leads to a high noise level in all multipole range and correlated features in the form of stripes due to the residual atmosphere residuals as shown in Figure \ref{fig:atm_sim_Q_maps}(b). In contrast, POMME with any value of $\tau > 1$ not only flattens the noise curve, recovering essentially white noise in the map domain, but simultaneously reduces the noise level. Note that having a smaller $\tau$ means that more degrees of freedom are lost during deprojection, resulting in a greater loss of information and hence greater noise levels. The noise level plateaus with a sufficiently large $\tau$ reaching the level as expected were the atmospheric contribution be absent from the onset, Fig.~\ref{fig:pomme_information_loss} (bottom panel). All these observations are consistent with our time-domain-based discussion in Section~\ref{sec:formalism}.

For our simulation setup, $ f_\mathrm{samp}/4f_\mathrm{HWP}\approx24.8$. We propose $\tau_{3/2}\equiv \lfloor 3 f_\mathrm{samp}/8f_\mathrm{HWP}\rfloor=37$ to be the optimal choice of $\tau$ following the arguments in Section \ref{sec:choosing_tau}. Indeed, we find that $\tau=37$ yields one of the smallest residual spectra in Figure \ref{fig:atm_auto_minus_cross_tau}, and outperforms $\tau_1=25$ overall.

\subsection{Quantifying I-P leakage}

\begin{figure*}
    \centering
    \includegraphics[width=\textwidth]
    {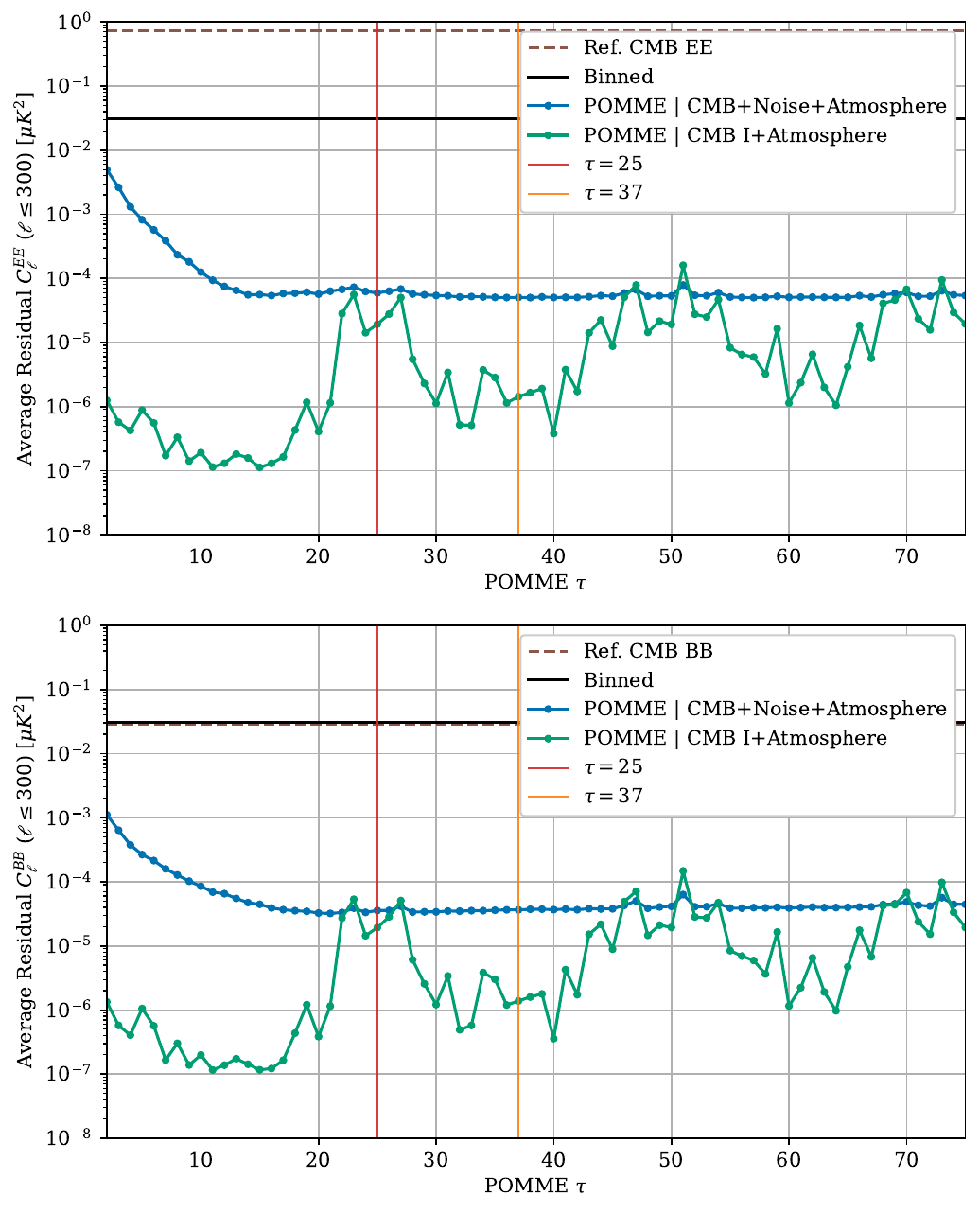}
    \caption{The average value of residual $EE$ and $BB$ angular power spectra obtained from binned mapmaker and POMME, for various values of $\tau$. The average is taken within the $\ell$ range of [10,300], with the corresponding values for the binned mapmaker estimate (black) and the underlying CMB map (red) shown as references. In addition to the POMME results on atmospheric simulations (blue), a run without any polarisation signals included (green) is shown to quantify the amount of I-P leakage. Overall, the proposed value of $\tau=1.5 f_\mathrm{samp}/4f_\mathrm{HWP}\approx 37$ controls the leakage level while maintaining a near-optimal noise level. }
    \label{fig:average_residuals_by_tau}
\end{figure*}

Figure \ref{fig:average_residuals_by_tau} shows the average level of residual $C_\ell$ on large scales for various values of POMME $\tau$ shown in Figure \ref{fig:atm_auto_minus_cross_tau}. The binned and POMME map estimates marked `CMB+Noise+Atmosphere' are the same as those in Figure \ref{fig:atm_auto_minus_cross_tau}. As before, smaller $\tau$ values increase the overall noise levels in the map, which leads to larger residual $C_\ell$s. The residuals are minimised at $\tau=38$ and $\tau=39$ for the $EE$ and $BB$ spectra, respectively. While $\tau_{3/2}=37$ is only $\sim2\%$ worse, $\tau_1=25$ has $\sim 40\%$ larger residuals in both cases.

We further test POMME on two datasets to quantify the amount of I-P leakage, i.e. contributions to polarisation from the unpolarised components of the data. For the dataset marked `CMB I+Atmosphere', only the CMB temperature and atmospheric emission have been included. The data contains neither polarised signals nor random noise at $f=4f_\mathrm{HWP}$, so the polarisation map should be as close to $0$ as possible. Here, we find that the I-P leakage remains orders of magnitude lower than the target signal. The suppression is particularly effective for small $\tau$. However, near $\tau=[mf_\mathrm{samp}/4f_\mathrm{HWP}]=25,50,\cdots$, the leakage level noticeably increases. This observation is consistent with the low-frequency leakage described in Section \ref{sec:frequency_response}; the deprojection transfers some low-frequency power into $4f_\mathrm{HWP}$ for these values of $\tau$ and increases the leakage. Concretely, the leakage level at $\tau_1=25$ is $20$ times larger than $\tau_{3/2}=37$. When only the CMB temperature data were used (`CMB I'), the leakage is much smaller in general but still larger at similar locations.

\begin{figure*}
    \centering
    \includegraphics
    {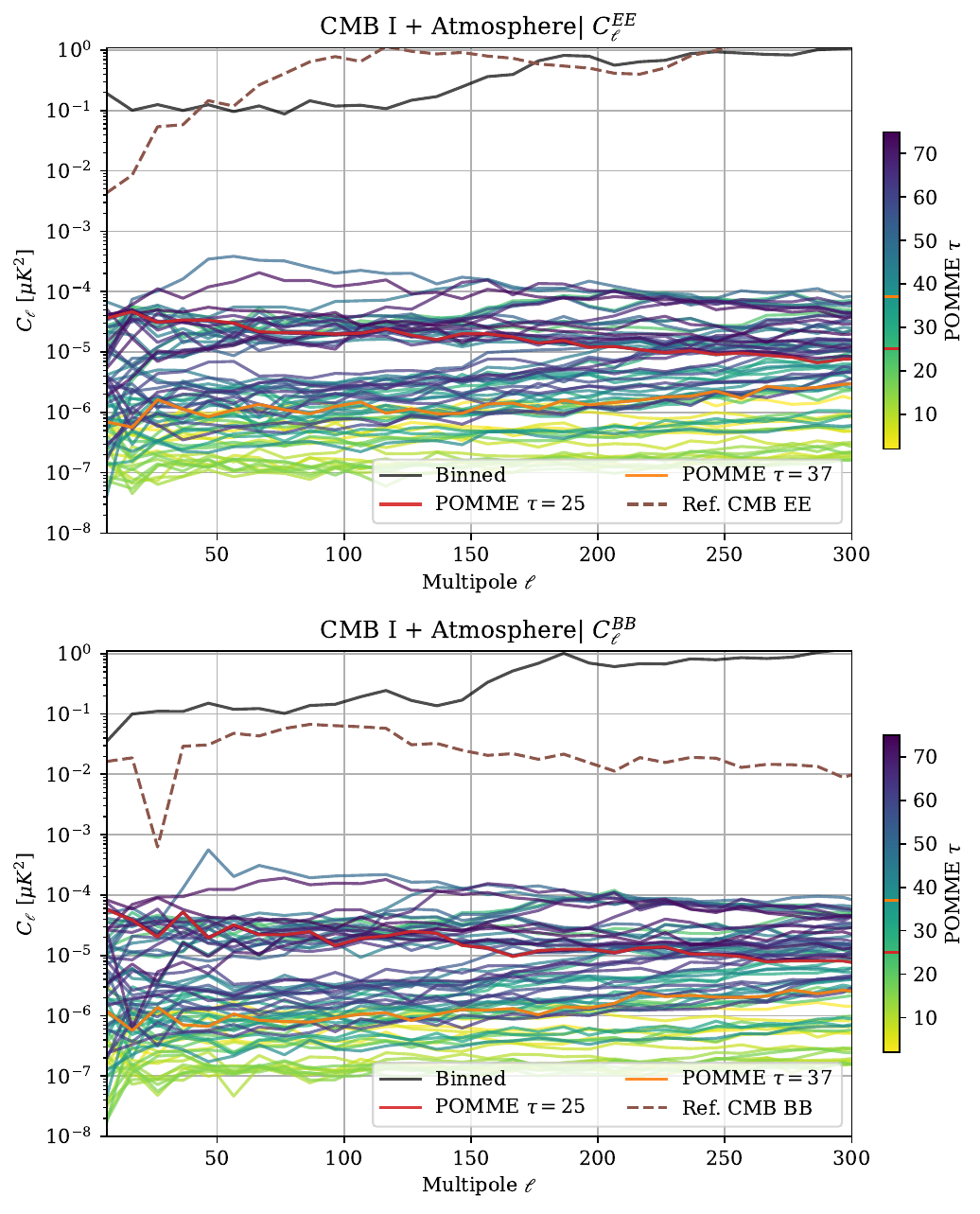}
    \caption{Noise levels on the polarisation power spectra analogous to Figure \ref{fig:atm_auto_minus_cross_tau}, but applied to TODs generated only with CMB temperature and atmospheric emission instead. No instrumental noise is present, but the noise in intensity partially leaks into polarisation at frequencies specified in Section \ref{sec:frequency_response}. Overall, the proposed value of $\tau= 1.5f_\mathrm{samp}/4f_\mathrm{HWP} \approx 37$ controls the leakage level at large scales, far below the target EE and BB spectra under study.
 }
    \label{fig:leakage_test}
\end{figure*}

The effect of I-P leakage on POMME map estimates is further detailed in Figure \ref{fig:leakage_test}. The residual $EE$ and $BB$ power spectra are plotted for the binned mapmaker and POMME with $2\le \tau \le 50$, using the dataset `CMB I + Atmosphere'. This covers a wider range of multipoles $\ell$ than Figure \ref{fig:atm_auto_minus_cross_tau}, and also includes the input CMB's power spectra for reference. Note that the input CMB decays quickly at small scales due to our relatively wide beam (27.4 arcmin). We make the following observations:

\begin{itemize}
    \item For all values of $\tau$, POMME effectively controls the I-P leakage level well below the input CMB in large scales $\ell \le 500$. In contrast, the naïve binned mapmaker is severely affected by the atmospheric emission, orders of magnitude larger than the polarised signal.
    \item The leakage level's dependence on $\tau$ at large scales is consistent with Figure \ref{fig:average_residuals_by_tau}: the lowest level at small $\tau$ and the highest around $\tau_1=25$ and $\tau_2=50$. Our proposed value of $\tau_{3/2}=37$ yields a leakage two orders of magnitude smaller than $\tau_1$.
\end{itemize}

\section{Conclusion} \label{sec:discussion}

The next generation of ground-based CMB polarisation experiments, including the Simons Observatory, will employ half-wave plates (HWPs) to modulate the polarisation signal. However, unpolarised signals such as atmospheric emission and ground pickup are often orders of magnitude larger than the target polarisation signal. Developing robust and computationally efficient mitigation strategies for these slowly varying contaminants is therefore critical for upcoming surveys.

In this work, we have introduced POMME (Polarisation-Optimised Map-Making Estimator), a novel template-based mapmaker designed to mitigate all slowly-varying signals. POMME operates by deprojecting constant values within intervals of fixed size $\tau$: an operation functionally analogous to a high-pass filter, but applied directly in the time-domain sample space. The method is computationally efficient and requires only a small number of preconditioned conjugate gradient (PCG) iterations. The loss of optimality in the map estimation is minimal despite the large number of templates used. POMME is implemented within the open-source \furax\ software package, leveraging the JAX framework for efficient, GPU-accelerated computation.

We validated POMME on a suite of simulated observations that incorporates realistic atmospheric emission. The results demonstrate that POMME successfully mitigates the atmospheric contamination $\mathcal O(10^3)$ times larger than the target polarisation signal. The output map noise is suppressed by a factor of $10^2-10^5$ in large angular scales compared to a simple binned mapmaker, with a white spectrum containing minimal residual imprint from the slowly varying atmospheric component.

POMME has one tunable parameter $\tau$: the number of samples in each interval used for template deprojection. We thoroughly investigated how $\tau$ affects POMME's overall performance and provided concrete analytic and numerical grounds for making the optimal choice: $\tau=\lfloor 3f_\mathrm{samp}/8f_\mathrm{HWP}\rfloor$. This is known \textit{a priori} as it depends only on the observation specifications\textemdash sampling rate and the HWP rotation frequency.

Several important directions remain for future investigation. First, while the simple boxcar template in POMME has proven effective, alternative options could be explored. For example, including low-order polynomials within each interval or using an overlapping set of intervals might be useful despite the increased computational cost. Second, the POMME formalism can be generalised to include other templates to mitigate high-frequency systematic effects such as the HWP synchronous signal. Third, a systematic benchmarking of POMME against alternative mapmaking approaches, notably``lock-in'' demodulation, as well as other filter-bin and maximum likelihood methods, will be essential to characterise the relative performance trade-offs.  Lastly, and most importantly, the method must be validated on real observational data from the Simons Observatory to assess its performance in the presence of non-idealities not captured by our simulation framework.

\begin{acknowledgments}

The authors would like to thank Hamza El Bouhargani, Michael Brown, Binh Nguyen, Benjamin Beringue and the members of the SCIPOL team for useful discussions during various stages of the project. We also thank Pierre Chanial for developing the \furax\ framework (\url{https://github.com/CMBSciPol/furax}), upon which the codebase for this project was built.

This work was carried out within the \textsc{SciPol} project (\url{https://scipol.in2p3.fr}), supported by the European Research Council (ERC) under the European Union’s Horizon 2020 research and innovation programme (Grant Agreement No.~101044073, PI: Josquin Errard). SB acknowledges support from the Science and Technology Facilities Council (grant number UKRI1164).

Computations were performed on the \textit{Jean Zay} supercomputer at IDRIS, using HPC resources provided by GENCI under allocation 2024-AD010414161R2 and 2025-A0190416919. Several authors acknowledge financial support from the Centre national d’études spatiales (CNES), France (ROR: \url{https://ror.org/04h1h0y33}), within the framework of the LiteBIRD space mission.

This work has also received funding by the European Union’s Horizon 2020 research and innovation program under grant agreement No. 101007633 CMB-Inflate.

\end{acknowledgments}

\appendix

\section{
Impact of the deprojection on the pixel noise.
}
\label{sec:appendix_frequency_dependence}

\begin{figure}
    \centering
    \includegraphics[width=\columnwidth]{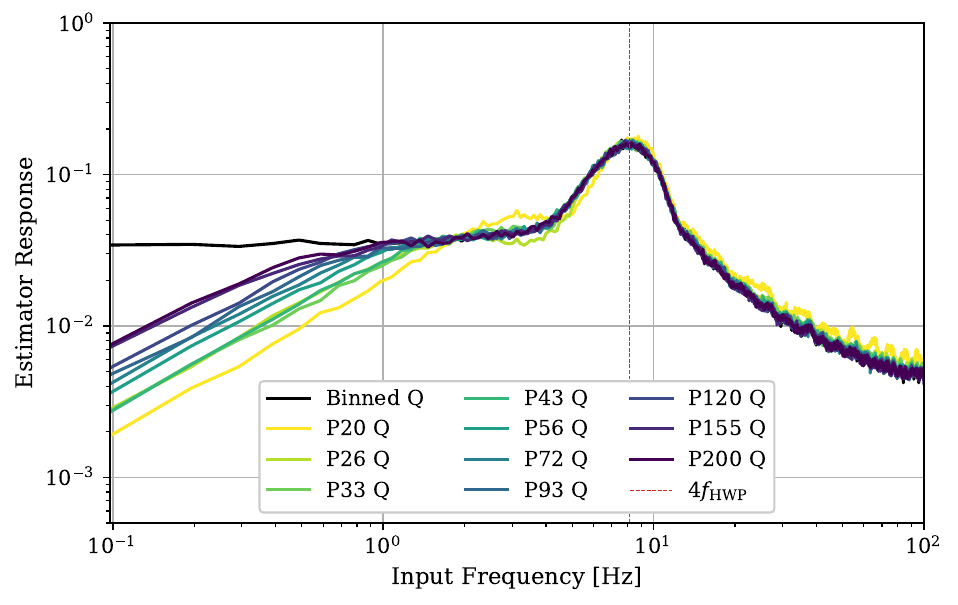}
    \caption{Response to random sinusoidal input for various mapmaker estimators: binned mapmaker (black) and POMME (Eq. \eqref{eqn:pomme_mapmaking_equation}, coloured, denoted P$(\tau)$). The response is computed for a single pixel with a typical scanning strategy detailed in the text and scales as $\propto n_\mathrm{hits}^{-1/2}$. POMME greatly reduces the polarisation map estimate's dependence on low-frequency noise, sometimes at the cost of increased high-frequency noise levels.}
    \label{fig:pomme_frequency_dependence}
\end{figure}

In section~\ref{sec:frequency_response} and Appendix~\ref{appendix:depNoisePS}, we elaborate on the effect of the POMME deprojector on the power spectrum of the time-domain noise. In this section, we complement this study by discussing the impact of the deprojection on the pixel-level noise.

For simplicity, we focus on a single sky pixel and assume a typical scanning strategy where the telescope performs a constant-elevation scan. The sky pixel is typically observed multiple times at various times by several detectors. The number of samples $n_\mathrm{crossing}$ corresponding to the pixel at each crossing is proportional to $f_\mathrm{sample}\theta_\mathrm{res}/\omega_\mathrm{scan}$, where the sky resolution $\theta_\mathrm{res}$ and scan speed $\omega_\mathrm{scan}$ are determined by our pixellisation and scanning strategy, respectively. In practice, $n_\mathrm{crossing}$ is roughly uniformly distributed between $1$ and $f_\mathrm{sample}\theta_\mathrm{res}/\omega_\mathrm{scan}$, since the detector pointing does not always pass through the centre of the pixel.

We study how a given input noise $\sin(2\pi f+\phi)$ with some frequency $f$ and random phase $\phi$ affects the polarisation map estimate. Since this noise is uncorrelated with the half-wave plate angle on average, an ideal mapmaker has a vanishing response to this signal. In reality, the response is non-zero and roughly scales as $\propto n_\mathrm{hits}^{-1/2}$, where $n_\mathrm{hits}$ is the number of samples corresponding to the given sky pixel.

Restricting our attention to a single pixel and the Stokes component Q, we can estimate the mapmaker's frequency response by taking the Fourier transform of the mapmaking operator: $(\mat{P}^\top \mat{N}^{-1}_\mathrm{diag} \mat{P})^{-1} \mat{P}^\top \mat{N}^{-1}_\mathrm{diag}$ for the binned case, and $(\mat{P}^\top \mat{N}^{-1} \mat{D}_\mat{T} \mat{P})^{-1} \mat{P}^\top \mat{N}^{-1} \mat{D}_\mat{T}$ for POMME. Note that the cross-pixel correlation is neglected in this particular exercise to speed up the computation. However, in principle, one can take the full multi-pixel operator and average over the pixels.

Figure \ref{fig:pomme_frequency_dependence} shows the result for various choices of $\tau$. We indeed confirm that the polarisation map estimate is less sensitive to low-frequency signals compared to the binned case, with a minimal increase in the high-frequency noise. The suppression is greater for smaller values of $\tau$. The binned mapmaker is equivalent to the limit $\tau\rightarrow\infty$.

\section{Deprojected noise power spectrum}

\label{appendix:depNoisePS}

Denoting by $\mathcal{F}$ the Fourier operator, the deprojected noise power spectrum is given by
\begin{align}
\mathcal{P}_{\mathrm{dep}}(f_i) \; = \; \big(\mathcal{F} & \,\mathcal{N}\,\mathcal{F}^\dagger\big)_{ii},
\end{align}
where the deprojected noise covariance, $\mathcal{N}$, is defined in Eq.~\eqref{eqn:depNoiseCov}. 
Using Eq.~\eqref{eqn:pomme_deproj_general} and assuming white noise weights, we can further write it as,
\begin{widetext}
\begin{align}
\mathcal{P}_{\mathrm{dep}}(f_i) \;&  = \; \big(\delta_{ij}\,-\,(\mathcal{F}\,\mat{T})_{it}\,(\mathcal{F}\,\mat{T})^\dagger_{jt}\big)\,\mathcal{P}(f_j)\,\delta_{jk}\, \big(\delta_{ki}\,-\,(\mathcal{F}\,\mat{T})_{ks}\,(\mathcal{F}\,\mat{T})^\dagger_{is}\big) \nonumber \\
& = \; \big(\delta_{ij}\,-\,(\mathcal{F}\,\mat{T})_{it}\,(\mathcal{F}\,\mat{T})_{jt}^\dagger\big)\,\mathcal{P}(f_j)\,
\big(\delta_{ji}\,-\,(\mathcal{F}\,\mat{T})_{js}\,(\mathcal{F}\,\mat{T})^\dagger_{is}\big) \nonumber \\
& = \; \mathcal{P}(f_i) \, -  \, 2\,\mathcal{P}(f_i) \, (\mathcal{F}\,\mat{T})_{it}\,(\mathcal{F}\,\mat{T})^\dagger_{it}
\, + \, (\mathcal{F}\,\mat{T})_{it}\,(\mathcal{F}\,\mat{T})^\dagger_{jt}\, \mathcal{P}(f_j)\, (\mathcal{F}\,\mat{T})_{js}\,(\mathcal{F}\,\mat{T})^\dagger_{is}.
\label{eq:depNoiseDiagStat}
\end{align}
\end{widetext}
Here, the subscripts $i, j, k$ denote the frequencies, and $t$ and $s$ the templates, and the summation convention is assumed for all repeated indices but $i$. The input noise, including all the contributions but the polarised sky signal, is assumed to be stationary and described by the noise power spectrum $\mathcal{P}(f)$. While this is clearly not always the case, e.g., for the scan synchronous or atmospheric signals, it is sufficiently general to allow us to gain some insight into the impact of the deprojection.

Given that all the templates in Eq.~\eqref{eqn:pomme_template} are merely time-shifted versions of each other, we can relate them to the 0th template by
\begin{align}
(\mathcal{F}\,\mat{T})_{it}\; = \; \exp(\iota\,2\pi f_i\,\tau\,t\,\Delta)\;(\mathcal{F}\,\mat{T})_{i0}.
\end{align}
where $\Delta \equiv T/n_{\mathrm{t}}$ is the sampling time and $f_i$ denotes the $i$th frequency, which we assume to be given here as $f_i = i/\Delta/n_{\mathrm{t}}; \ i = -n_{\mathrm{t}}/2+1, \dots, n_{\mathrm{t}}/2$. We then can rewrite Eq.~(\ref{eq:depNoiseDiagStat}) as,
\begin{widetext}
\begin{align}
\big(\mathcal{F} & \,\mathcal{N}\,\mathcal{F}^\dagger\big)_{ii} \;   = 
\;  \mathcal{P}(f_i) \left(1 \, -  \, \genfrac{}{}{}{0}{n_{\mathrm{t}}}{\tau} \, |(\mathcal{F}\,\mat{T})_{i0}|^2\right)^2 \, + \, \left(\genfrac{}{}{}{0}{n_{\mathrm{t}}}{\tau}\right)^2 \, |(\mathcal{F}\, \mat{T})_{i0}|^2\,\sum_{\genfrac{}{}{0pt}{1}{k=-\Big\lfloor \genfrac{}{}{}{1}{n_{\mathrm{t}}/2+i-1}{\lfloor n_{\mathrm{t}}/\tau\rfloor}\Big\rfloor}{k\neq 0}}^{\Big\lfloor\genfrac{}{}{}{1}{n_{\mathrm{t}}/2-i}{\lfloor n_{\mathrm{t}}/\tau\rfloor}\Big\rfloor} \,|(\mathcal{F}\,\mat{T})_{i+k \lfloor n_{\mathrm{t}}/\tau\rfloor, 0}|^2\,  \mathcal{P}(f_{i+k\lfloor n_{\mathrm{t}}/\tau\rfloor}),
\label{eqn:depPS}
\end{align}
\end{widetext}
where $\lfloor\cdot \rfloor$ denotes the integer part of a number.

For the boxcar templates assumed here, we can explicitly calculate the Fourier transforms of the templates. Inserting them in Eq.~\eqref{eqn:depPS} yields Eq.~\eqref{eqn:genDepNoisePS}.

\bibliography{references}

\end{document}